\documentclass[aps,showpacs,twocolumn,prb,amsmath,amssymb,superscriptaddress,intlimits,longbibliography,showkeys]{revtex4-1}

\usepackage{graphicx}
\usepackage{subfigure}
\usepackage{xcolor}
\usepackage{epstopdf}
\usepackage{amssymb}
\usepackage{amsmath}
\usepackage{amsfonts}
\usepackage{dcolumn}
\usepackage{bm}
\usepackage{soul}
\usepackage{array}
\usepackage{physics}

\newcommand{\vk}{{\bf k}}
\newcommand{\vQ}{{\bf Q}}
\newcommand{\vq}{{\bf q}}

\newcolumntype{C}[1]{>{\centering\let\newline\\\arraybackslash\hspace{0pt}}m{#1}}

\graphicspath{{figures/}}

\begin{document}

\title{Exciton-defect interaction and optical properties from a first-principles T-matrix approach}

\author{Yang-hao Chan}
\email{yanghao@gate.sinica.edu.tw}
\affiliation{Institute of Atomic and Molecular Sciences, Academia Sinica, Taipei 10617, Taiwan}

\author{Jonah B. Haber}
\affiliation{Department of Materials Science and Engineering, Stanford University, Stanford, CA 94305, USA}

\author{Mit H. Naik}
\affiliation{Department of Physics, University of Texas at Austin, Austin, TX, 78712, USA}

\author{Diana Y. Qiu}
\email{diana.qiu@yale.edu}
\affiliation{Department of Materials Science, Yale University, New Haven, CT 06520}

\author{Felipe H. da Jornada}
\email{jornada@stanford.edu}
\affiliation{Department of Materials Science and Engineering, Stanford University, Stanford, CA 94305, USA}

\begin{abstract}
Understanding exciton-defect interactions is critical for optimizing optoelectronic and quantum information applications in many materials. However, \textit{ab initio} simulations of material properties with defects are often limited to high defect density. Here, we study effects of exciton-defect interactions on optical absorption and photoluminescence spectra in monolayer MoS$_2$ using a first-principles T-matrix approach.
We demonstrate that exciton-defect bound states can be captured by the disorder-averaged Green's function with the T-matrix approximation and further analyze their optical properties.
Our approach yields photoluminescence spectra in good agreement with experiments and provides a new, computationally efficient framework for simulating optical properties of disordered 2D materials from first-principles.

\end{abstract}

\date{\today}
\maketitle

\textit{Introduction} Optical properties of materials can be strongly affected by the presence of defects through changes of energy levels and introduction of new in-gap states~\cite{Tongay2013,Chow2015,Hong2015,Lin2016,Khan2017,Klein2019,Mitterreiter2021}. In quasi-two dimensional systems, strongly bound excitons (correlated electron-hole pairs) dominate the low-energy optical spectra. A thorough understanding of exciton-defect interactions is critical to the fundamental understanding of exciton character and dynamics in disordered systems~\cite{Rosenberger2018,Greben2020,Wu2022} and is essential for optoelectronic devices and quantum information applications~\cite{Koperski2015,He2015,Srivastava2015,Chakraborty2015,ChakrabortyVamivakasEnglund2019,Barthelmi2020}.

State-of-the-art calculations of optical properties with defects, utilizing a supercell approach, have been performed at the GW plus Bethe-Salpeter equation (BSE) level, where both quasi-particle energy renormalization and electron-hole interactions are taken into account~\cite{Attaccalite2011,Sivan2018,Mitterreiter2021}. 
However, due to the large computational cost associated with calculating excited states in large supercells, only systems with high defect density, where defect-defect interactions and artificial periodicity might obscure single-defect properties, have been studied. A thorough investigation of the defect density dependence of the optical spectrum, especially in low-density experimentally relevant regimes, has not yet been conducted. 

One way of accessing the dilute defect limit is to treat the defect potential as a perturbation acting on the excitonic states of the pristine system: in this picture, the exciton propagates through the host material and is scattered by the localized potential introduced by the defect. The lowest order treatment of this interaction---the Born approximation---accounts for single scattering events but cannot capture defect-bound states, where it is necessary to coherently resum scattering events to infinite order. This infinite resummation, encapsulated by the T-matrix formalism, allows poles to build in the exciton Green's function, corresponding to defect-bound states.

In this paper, we develop an efficient \textit{ab initio} T-matrix approach to study exciton-defect interactions applicable to a wide range of defect densities. 
We find that, in addition to the A and B peaks of MoS$_2$, a weak absorption peak appears when defects are included, which can be identified as a defect-bound state. We benchmark two common methods to incorporate disorder-averaged self-energies, and show that, unlike the T-matrix, the Born approximation is incapable of qualitatively capturing defect-bound excitons. We compute the photoluminescence (PL) spectra including defect scattering from first principles and observe clear signals from defect-bound states that completely dominate the PL spectrum at low temperatures, despite their lower oscillator strengths, in good agreement with available experiments.
We anticipate that this framework will open new avenues for investigation into exciton-defect interaction, enabling systematic studies across a range of defect densities at minimal computational cost---ultimately laying the groundwork for understanding even more complex processes, e.g. inelastic exciton-defect scattering.

\begin{figure}[t]
     \centering
     \includegraphics[width=0.45\textwidth]{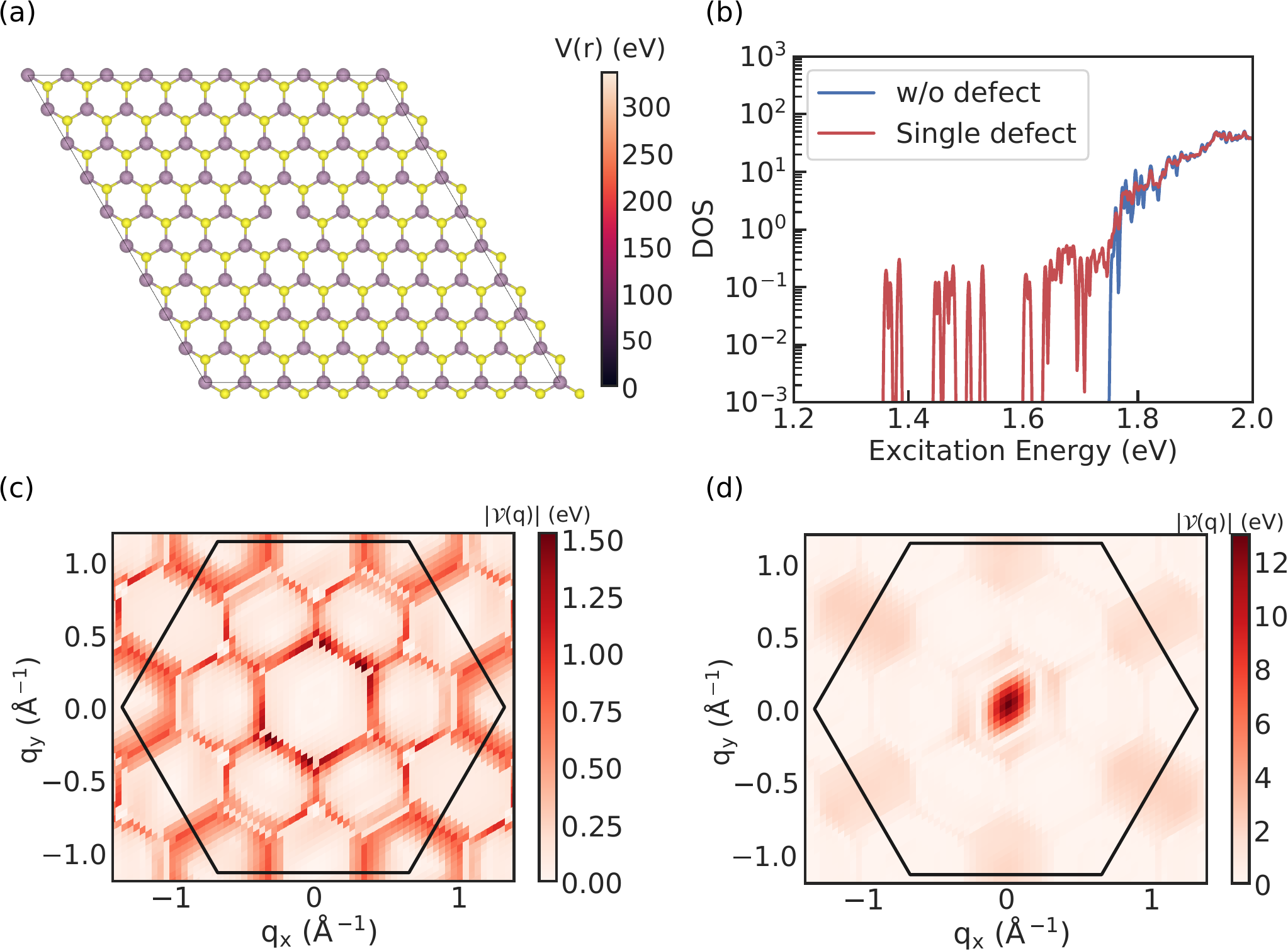}
\caption{ (a) Intensity map of a single S-vacancy defect potential in a $9\times9$ supercell overlaid with the atomic model, where purple balls are Mo atoms and yellow balls are S atoms. (b) Density of exciton states without (blue) and with (red) the single S-vacancy defect in the supercell. Absolute value of defect scattering matrix elements between A exciton and excitons (c) in the first and (d) the third exciton band in the Brillouin zone.}
\label{fig:single_defect}
\end{figure}

\textit{Method} We start by writing a Hamiltonian, defined in the primitive unit cell of a material, describing a set of excitons that can scatter with defects,
\begin{align}
    H=\sum_{S,\mathbf{Q}}E_{S\mathbf{Q}}a^\dagger_{S\mathbf{Q}}a_{S\mathbf{Q}}+\sum_{SS'\mathbf{Q}\mathbf{q}}V_{S\vQ+\vq,S'\vQ}a^\dagger_{S\mathbf{Q+q}}a_{S'\mathbf{Q}},
    \label{eq:Ham}
\end{align}
where $a_{S\mathbf{Q}}$ ($a^\dagger_{S\mathbf{Q}}$) are exciton annihilation (creation) operators for an exciton state $S$ with finite center of mass momentum (COM) $\mathbf{Q}$ and $V_{S\vQ+\vq,S'\vQ}$ describes the scattering amplitude between exciton state $(S',\vQ)$ and $(S,\vQ+\vq)$. 
In this work, we consider the sulfur-vacancy (S-vacancy) defect, which is commonly observed in MoS$_2$~\cite{Greben2020,Barthelmi2020,Mitterreiter2021}. The defect potential is extracted from density-functional theory (DFT) calculations and requires no explicit knowledge of Kohn-Sham states in large supercells ~\cite{Perturbo,Lu2019,Lu2020,Kristen2020,Xu2021}. The single S-vacancy defect potential is shown in Fig.~\ref{fig:single_defect} (a) for a $9\times9\times1$ supercell and the resulting exciton density of states of Eq.~(1) is shown in Fig.~\ref{fig:single_defect} (b). The exciton-defect scattering matrix element is calculated as 
\begin{widetext}
\begin{equation}
    V_{S\vQ+\vq,S'\vQ}=\sum_\mathbf{k}\left[\sum_{vcc'}A^{S*}_{c\mathbf{k}+\mathbf{Q}+\mathbf{q},v\mathbf{k}}A^{S'}_{c'\mathbf{k+Q},v\mathbf{k}}V_{cc'}(\mathbf{k}+\mathbf{Q},\mathbf{q})-\sum_{cvv'}A^{S*}_{c\mathbf{k+Q},v\mathbf{k-q}}A^{S'}_{c\mathbf{k+Q},v'\mathbf{k}}V_{v'v}(\mathbf{k-q,q})\right],
    \label{eq:Vss}
\end{equation}
\end{widetext}
where $A^S_{cv\vk}$ is the exciton envelope function obtained by solving the BSE~\cite{Deslippe2012,Hybertsen1986,Rohlfing2000} and $V_{nm}(\vk,\vq)$ is the electron-defect matrix element between electron states $(m,\vk)$ and $(n,\vk+\vq)$. The first (second) term in Eq.~\ref{eq:Vss} describes electrons (holes) scattering off defects. The details of the calculations are given in the supplementary materials (SM)~\cite{SM}.

We show in Fig.~\ref{fig:single_defect} (c) and (d) the exciton-defect matrix elements between the A exciton, with $\mathbf{Q=0}$, and other excitons with COM $\vq$ in the first and third exciton bands, respectively. We observe that the momentum distribution of the matrix elements slightly breaks the three-fold rotation symmetry, which is due to structural relaxation in our defect simulations. 
The vanishing matrix elements near $\vq=0$ in Fig.~\ref{fig:single_defect} (c) and the large amplitude in (d) can be understood from spin quantum number conservation. Since the S-vacancy does not induce spin-flip processes and spin is almost a good quantum number for low energy excitons in monolayer MoS$_2$~\cite{Cao2012,Chan2023}, the A exciton can only scatter into parallel-spin states.

\begin{figure}[t]
     \centering
     \includegraphics[width=0.48\textwidth]{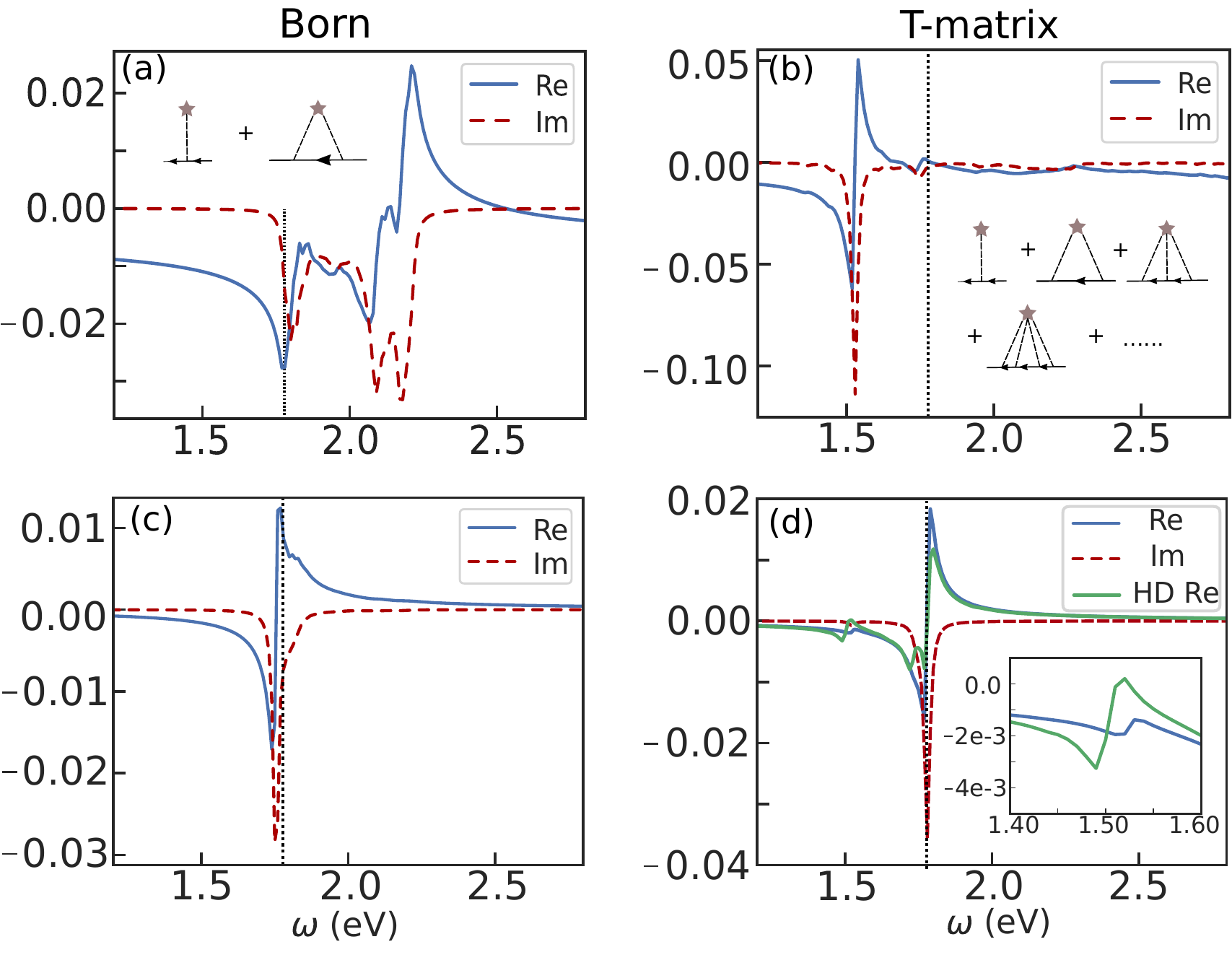}
\caption{Retarded (a) Born and (b) T-matrix self-energy of A exciton calculated with a defect density of $5\times10^{11}\,$cm$^{-2}$. (c) and (d) show the corresponding Green's functions. Blue solid lines (red dashed lines) are the real (imaginary) part. Black dotted lines mark the bare A exciton energy. In (d), the real part of the Green's function with a defect density of $2.5\times10^{12}\,$cm$^{-2}$ is shown in green solid line. Insets in (a) and (b) show the self-energy diagrams in each approximation. The inset in (d) is a zoom-in view around 1.5 eV.}
\label{fig:SEandG}
\end{figure}

\textit{Born and T-matrix approximations} Up to this point, we have focused on the single defect-per-supercell problem, which can be solved by exact diagonalization. However, one would like to access various---possibly dilute---defect concentrations and avoid explicitly diagonalizing Eq.~\ref{eq:Ham}~\cite{Bruus2004}, which we achieve within our Green's function approach through a disorder-averaging procedure~\cite{Bruus2004,Kristen2020}. 
Importantly, such a defect-averaging process recovers the translational symmetry of the primitive unit cell of the material---broken by an isolated defect or array of defects in a supercell---and leads to a self-energy diagonal in the exciton's COM crystal momentum. 
In the following, we adopt Born and T-matrix self-energy approximations and compare their self-energy, Green's function, and the absorption spectrum.
The self-energy from Born approximation describes scattering processes shown in the inset of Fig.~\ref{fig:SEandG} (a). In contrast, the quasiparticle can scatter off defects multiple times in the T-matrix approximation as illustrated in the inset of Fig.~\ref{fig:SEandG} (b). While the Born self-energy is the lowest order non-trivial one, it has been shown that the T-matrix approximation becomes exact in the low defect density limit~\cite{Bruus2004}. 

The Born self-energy is written as
\begin{align}
   \Sigma^B_{SS'\vQ}(\omega) &= N_i \big[ V_{S\vQ,S'\vQ} \nonumber\\&+\sum_{S''\vQ'}V_{S\vQ,S''\vQ'}G^0_{S''\vQ'}(\omega)V_{S''\vQ',S'\vQ} \big], 
\end{align}
where $N_i$ is the number of defects, $G^0_{S\vQ}=1/(\omega-E_{S\vQ}+i\eta)$ is the retarded bare exciton Green's function with $\eta=10$ meV in our calculations.
For the T-matrix self-energy, we have $\Sigma^T_{\vQ}(\omega)=N_iT_{\vQ\vQ}(\omega)$ with ~\cite{Bruus2004,Kristen2020}
\begin{align}
    T_{S\vQ S'\vQ'}(\omega)&=V_{S\vQ S'\vQ'}+\sum_{S''Q''}\big[V_{S\vQ S''\vQ''}\nonumber\\
    &\times G^0_{S''\vQ''} (\omega)T_{S''\vQ''S'\vQ'}(\omega)\big].
\end{align}
The full Green's function is solved from Dyson's equation,
$G(\omega)^{-1} = G^{0}(\omega)^{-1} - \Sigma(\omega)$.

In Fig.~\ref{fig:SEandG}, we show the self-energy and the Green's function in the two approximations for the A exciton with a defect density of $5\times10^{11}\,$cm$^{-2}$. Notably, the Born self-energy is featureless below the bare exciton energy, defined as that in the material without defects, while the T-matrix self-energy acquires a lower-energy pole at 1.53 eV, which can be assigned as the defect-bound state and will be denoted as ``Bd1'' in the following~\cite{SM}. 
As a result, a corresponding secondary peak appears in the T-matrix Green's function. This striking difference between the Born and T-matrix self-energy is well-documented: because the Born approximation accounts for a finite (two) scattering events between excitons and defects, it cannot capture bound states. At most, it describes a renormalization of the exciton energies due to the change of the average potential they experience. In contrast, the T-matrix allows for an infinite number of scatering events between excitons and defects, allowing for a bound state to emerge.

As the defect density increases, the Born and T-matrix approximations also display contrasting behaviors. In the Born approximation, the renormalization of the exciton energy increases with the defect density, without the appearance of any lower-energy peak associated with a defect-bound exciton. In contrast, for the T-matrix, as the defect density increases, more spectral weight gets transfered to the lower-energy secondary peak, with little renormalization of the bare exciton energies. The T-matrix approximation hence correctly captures the physics that there are more defect-bound excitons in highly disordered samples, while the Born approximation is qualitatively incapable of describing excitons in defected materials.

\begin{figure}[t]
     \centering
     \includegraphics[width=0.48\textwidth]{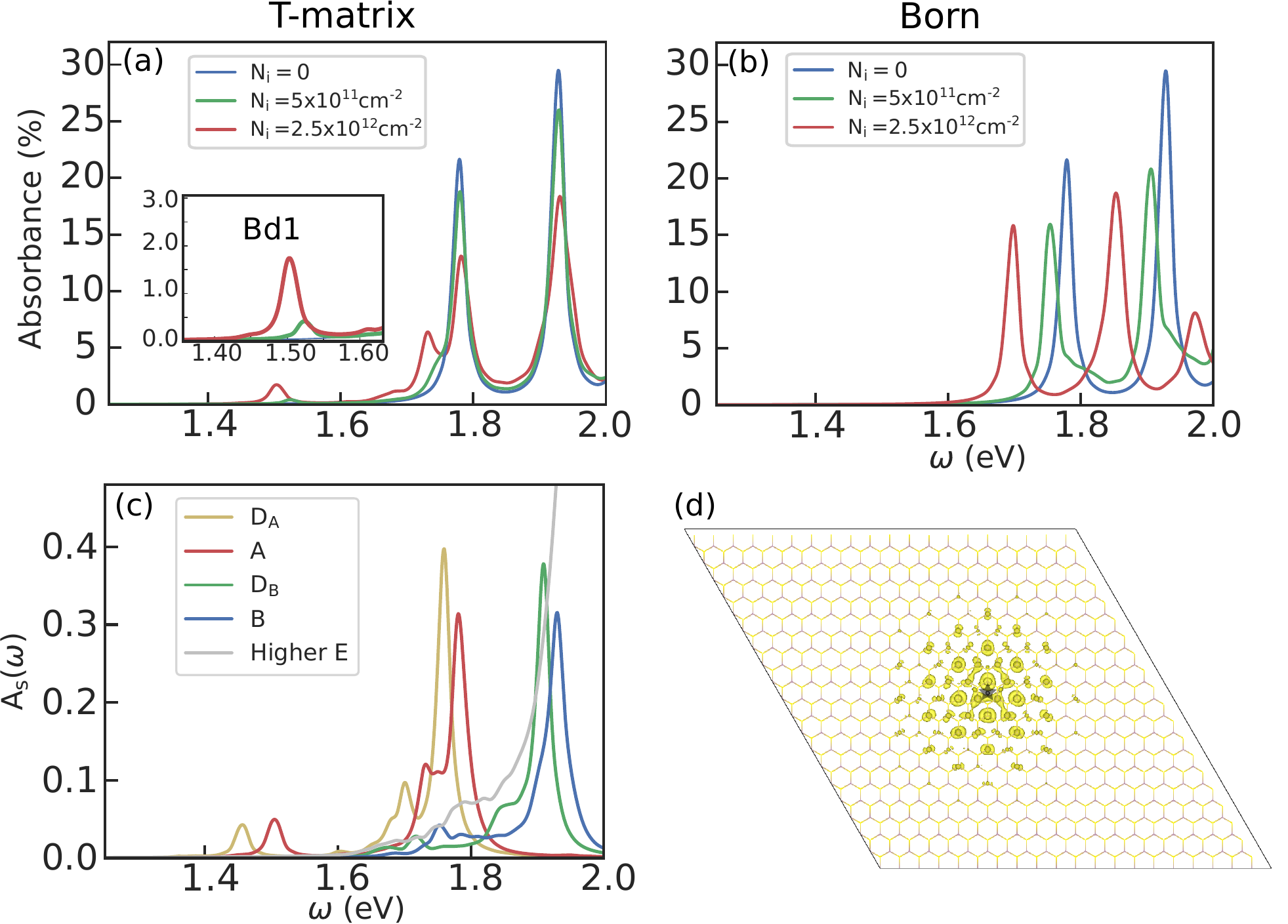}
\caption{Computed optical absorbance for monolayer MoS$_2$ within the (a) T-matrix and (b) Born approximations at several defect densities at 300 K. The inset in (a) shows a zoom-in view near the 1.5 eV. (c) Exciton-state decomposed spectral function $A(\omega)$ from T-matrix calculations for a defect density of $2.5\times10^{12}\,$cm$^{-2}$, which shows the contributions from the bare exciton states. $D_A$ and $D_B$ refers to the lowest-energy dark states in the same series as A and B excitons, respectively. (d) Local electron density of states of the lowest energy peak in (a) for a fixed hole position marked by a star symbol. }
\label{fig:abs}
\end{figure}

\textit{Absorption spectrum} The absorption spectrum including the exciton-defect interactions can be calculated from the retarded Green's function,
   $ \epsilon_2(\omega)=-\frac{e^2}{\epsilon_0V_{tot}}\text{Im}\sum_{SS'}\Omega^*_{S'}G^R_{S'S\vQ=0}(\omega)\Omega_{S},$
where $V_{tot}$ is the crystal volume, $\Omega_S=\sum_{cv\vk}A^{S*}_{cv\vk}d_{cv\vk}$ are exciton dipole matrix elements, and $d_{vc\vk}$ are electron dipole matrix elements.

In Fig.~\ref{fig:abs} (a) and (b) we show the spectra from T-matrix and Born approximations, respectively, at two defect densities and for the pristine case.
As expected from the Green's function calculations, we observe that A and B peaks shift to lower energy in Born approximation due to spurious exciton energy renomalizations with the average defect potential. 
In contrast, within the T-matrix approximation, their energy renomalizations are minimal. 
We instead observe the appearance of secondary peaks and shoulder structures, which can be identified as defect-bound states, and suppression of absorbance of A and B excitons.
We also find that the Bd1 peak shifts to 1.5 eV at a defect density of $2.5\times10^{12}\,$cm$^{-2}$, following the defect density dependence of the spectral function discussed earlier.
To understand how defect-bound states acquire oscillator strength and their compositions of bare excitons, we write the full Green's function as 
\begin{align}
    G^R_{SS'}(\omega)=\sum_\lambda\frac{T_{\lambda S}(\omega)T^{-1}_{\lambda S'}(\omega)}{\hbar\omega-\tilde{E}_\lambda(\omega)-i\tilde{\Gamma}_\lambda(\omega)},
    \label{eq:fullGR}
\end{align}
where $\tilde{E}_\lambda(\omega)-i\tilde{\Gamma}_\lambda(\omega)$ and $T_{\lambda S}(\omega)$ are eigenvalues and eigenvectors, respectively, of the effective Hamiltonian $H_{SS'}(\omega)=E_{S\mathbf0}\delta_{SS'}+\Sigma_{SS'}(\omega)$.
In Fig.~\ref{fig:abs} (c), we plot $A_S(\omega)=-\frac{1}{\pi}\text{Im}G^R_{SS}(\omega)$ for selected exciton states at a defect density of $2.5\times10^{12}\,$ cm$^{-2}$. Compared with Fig.~\ref{fig:abs} (a), we find that the Bd1 peak has an origin of bare A exciton states---a fact that is also reflected in similar real-space distribution of the electron in the defect-bound exciton and A exciton shown in Fig.~\ref{fig:abs} (d) (see SI)---while the 1.75 eV peak has contributions from both A and B excitons. 
In addition to bright states, we find a lower-energy peak at 1.45 eV originating from dark, spin-unlike excitons ($D_A$) in the pristine system, so they do not appear in the absorption spectrum. These states, and others extending up to $\sim$ 1.6~eV, can be connected to the low-energy eigenstates of Eq.~\ref{eq:Ham}. A general expression of $A(\omega)$ in terms of the eigenstates of Eq.~\ref{eq:Ham} is given in SM~\cite{SM}, where a clear connection to bound states of the single defect problem can be made. For higher-energy excitons, we do not find associated defect-bound states, but their spectra extend a few hundred meV below the main peaks.

\textit{PL spectrum} The effect of defects on the PL spectrum has been carefully studied in several experiments~\cite{Tongay2013,Chow2015,Rosenberger2018,Barthelmi2020,Mitterreiter2021} but has only been investigated from first-principles in a few works~\cite{Attaccalite2011,Sivan2018}.
In contrast to the absorption, PL intensity is proportional to the lesser Green's function~\cite{Hannewald2000,Marini2024} as
   $ I_{PL}(\omega)\varpropto\sum_S\left|\Omega_S\right|^2G^<_{SS\mathbf{0}}(\omega).$
In general, $G^<_{SS\vQ}(\omega)$ can be obtained from the Kubo-Martin-Schwinger relation, where $G^<(\omega)=b(\omega)(G^R(\omega)-G^A(\omega))$ with $b(\omega)$ being the Bose distribution function. 
However, it is difficult to deal with the Bose function numerically when $\text{Im} GR$ has a nontrivial structure, such as multiple sattelite peaks. 
Here, we calculate $G^<_{SS}(\omega)$ within a quasiparticle expansion~\cite{Bechstedt1994,Cudazzo2023,Marini2024} which separates the quasiparticle part and the dynamical contribution as
\begin{align}
    G^<_{SS\vQ}(\omega)&=-2\pi ib(\omega)\left[\delta(\omega-E_{S\vQ})(1-R_{S\vQ})\right.\nonumber\\
    &\left.+\frac{N_i}{(\omega-E_{S\vQ})^2+\eta^2}\sum_\lambda |W_{S\vQ}^\lambda|^2\delta(\omega-E^\lambda)\right],
    \label{eq:Glesser}
\end{align}
where $R_{S\vQ}$ is the renormalization factor, $E^\lambda$ is the eigenenergy of Eq.~\ref{eq:Ham}, and $W^{\lambda}_{S\vQ}$ is the exciton-defect matrix elements projected to the eigenvector of Eq.~\ref{eq:Ham}~\cite{SM}. The first term in Eq.~\ref{eq:Glesser} describes the renormalized quasiparticle peak with a reduced weight, while the second term is responsible for structures such as sattellites, obtained from dynamical effects. In this form of expressing $G^<$, the positions of the secondary peaks are solely determined by the energy of the discrete, defect-bound excitons from Eq.~\ref{eq:Ham} that is solved for a fixed defect density. Therefore, we can approximately associate those peaks obtained at various defect densities with defect-bound states shown in Fig.~\ref{fig:single_defect} (b).

\begin{figure}[t]
     \centering
     \includegraphics[width=0.49\textwidth]{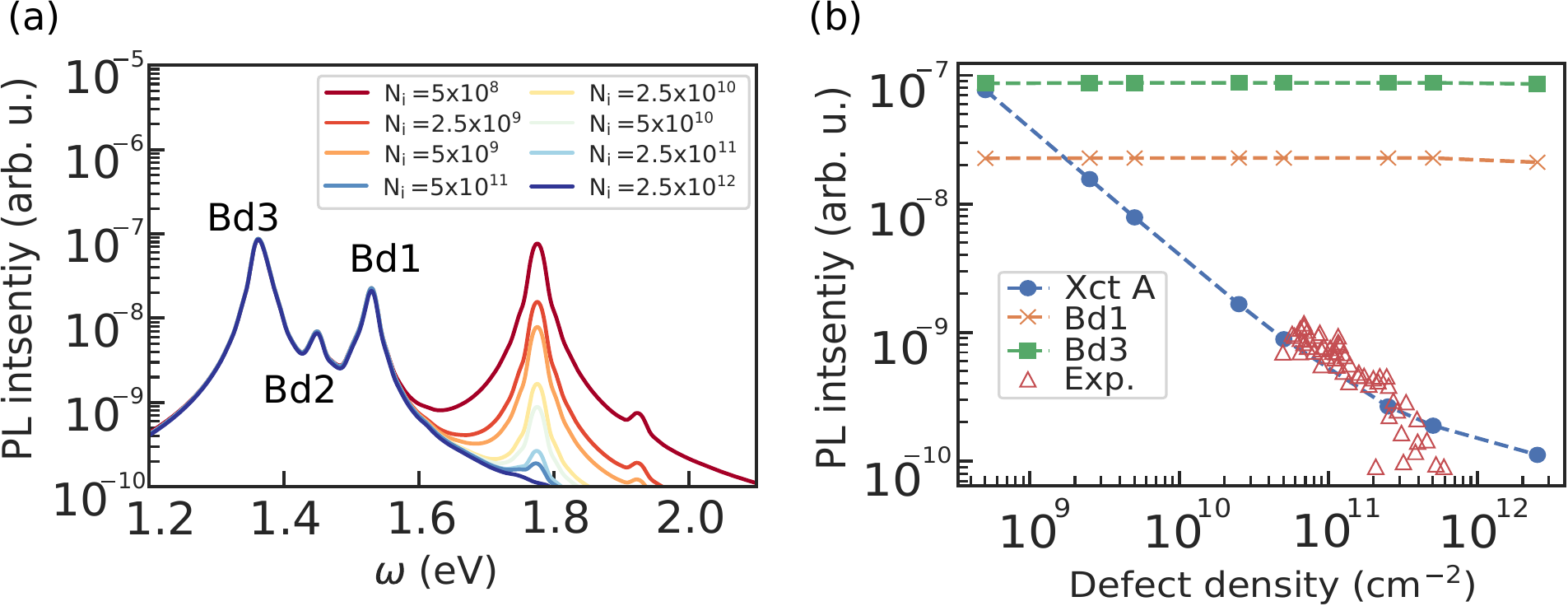}
\caption{Simulated photoluminescence spectra at (a) different defect density at 300 K. Defect densities are in the unit of cm$^{-2}$. (b) Defect density dependences of the PL intensity at A exciton (blue dots), Bd1 (orange crosses), and Bd3 (green squares) peaks. Experimental data in (b) is adopted from Ref.~\cite{Rosenberger2018} and scaled by an overall factor to align with the calculations.}
\label{fig:PL}
\end{figure}

In Fig.~\ref{fig:PL} (a), we show the simulated PL intensity at 300 K for several defect densities, where the spectra are normalized by the total number of excitons.
Besides the A, B, and Bd1 peaks at 1.78, 1.93, and 1.53 eV, we identified two additional defect-bound state emissions at 1.45 and 1.36 eV, denoted as ``Bd2'' and``Bd3'', respectively.
We find that the Bd3 state consists of both A exciton and the lowest energy $Q=0$ dark exciton, with the latter having about 4 times larger weight.
Bd1 and Bd2 states are both derived from the A exciton.
Notably, the detunings of the Bd1 and Bd2 peak with respect to the A peak are consistent with those observed in Ref.~\cite{Barthelmi2020}, where emissions with detuning of 195 meV and 275 meV were reported. 
The lack of clear experimental evidence for the Bd3 peak could be related to its weak oscillator strength: it implies that its emission is only possible if the system reaches thermal equilibrium and there are no other decay and scattering mechanisms faster than the Bd3 radiative recombination time.

At low defect density, the A peak intensity is comparable to the intensity of the defect states. 
With increasing defect density, A peak emission decreases rapidly, and the emission from the two bound states dominates the spectrum, since the PL intensity is directly proportional to exciton populations.
The first term in Eq.~\ref{eq:Glesser} is only weakly dependent on defect densities while the second term is proportional to it. Therefore, the A peak intensity must be inversely proportional to the defect density after normalization by the total number of excitons for each curve, as is indeed seen in Fig.~\ref{fig:PL} (b). Our results for monolayer MoS$_2$ are in remarkable agreement with available experimental data for monolayer WS$_2$~\cite{Rosenberger2018}. We attribute this agreement to similarities in the composition and electronic structure of these two closely related TMDs.

In conclusion, we develop a first-principles T-matrix approach for exciton-defect problems and apply it for S-vacancy in monolayer MoS$_2$, where optical absorption and PL intensity spectra are simulated. 
We identified defect-bound states in both spectra and revealed their characters in terms of excitons in pristine MoS$_2$.
Compared against the Born approximation, we found a T-matrix approach is necessary to capture defect-bound states because the former merely shifts the exciton energy.
The defect density dependence of the PL intensity and defect-bound state energy agrees reasonably well with the experiments.
Our approach can generally be applied to other materials and different types of defects. We anticipate that our approach can provide an understanding of exciton-defect couplings and the defect-bound exciton emission that complement the conventional supercell method limited to the high defect density. 

\section*{Acknowledgments}
\label{sec:acknowledgments}

This work was primarily supported by the Center for Computational Study of Excited State Phenomena in Energy Materials (C2SEPEM), which is funded by the U.S. Department of Energy, Office of Science, Basic Energy Sciences, Materials Sciences and Engineering Division under Contract No. DE-AC02-05CH11231, as part of the Computational Materials Sciences Program.
We acknowledge the use of computational resources at the National Energy Research Scientific Computing Center (NERSC), a DOE Office of Science User Facility supported by the Office of Science of the U.S. Department of Energy under Contract No. DE-AC02-05CH11231. The authors acknowledge the Texas Advanced Computing Center (TACC) at The University of Texas at Austin for providing HPC resources that have contributed to the research results reported within this paper. YHC thanks the National Center for High-Performance Computing (NCHC) for providing computational and storage resources.

\section*{References}
\bibliography{ref}

\begin{thebibliography}{34}%
\makeatletter
\providecommand \@ifxundefined [1]{%
 \@ifx{#1\undefined}
}%
\providecommand \@ifnum [1]{%
 \ifnum #1\expandafter \@firstoftwo
 \else \expandafter \@secondoftwo
 \fi
}%
\providecommand \@ifx [1]{%
 \ifx #1\expandafter \@firstoftwo
 \else \expandafter \@secondoftwo
 \fi
}%
\providecommand \natexlab [1]{#1}%
\providecommand \enquote  [1]{``#1''}%
\providecommand \bibnamefont  [1]{#1}%
\providecommand \bibfnamefont [1]{#1}%
\providecommand \citenamefont [1]{#1}%
\providecommand \href@noop [0]{\@secondoftwo}%
\providecommand \href [0]{\begingroup \@sanitize@url \@href}%
\providecommand \@href[1]{\@@startlink{#1}\@@href}%
\providecommand \@@href[1]{\endgroup#1\@@endlink}%
\providecommand \@sanitize@url [0]{\catcode `\\12\catcode `\$12\catcode
  `\&12\catcode `\#12\catcode `\^12\catcode `\_12\catcode `\%12\relax}%
\providecommand \@@startlink[1]{}%
\providecommand \@@endlink[0]{}%
\providecommand \url  [0]{\begingroup\@sanitize@url \@url }%
\providecommand \@url [1]{\endgroup\@href {#1}{\urlprefix }}%
\providecommand \urlprefix  [0]{URL }%
\providecommand \Eprint [0]{\href }%
\providecommand \doibase [0]{http://dx.doi.org/}%
\providecommand \selectlanguage [0]{\@gobble}%
\providecommand \bibinfo  [0]{\@secondoftwo}%
\providecommand \bibfield  [0]{\@secondoftwo}%
\providecommand \translation [1]{[#1]}%
\providecommand \BibitemOpen [0]{}%
\providecommand \bibitemStop [0]{}%
\providecommand \bibitemNoStop [0]{.\EOS\space}%
\providecommand \EOS [0]{\spacefactor3000\relax}%
\providecommand \BibitemShut  [1]{\csname bibitem#1\endcsname}%
\let\auto@bib@innerbib\@empty
\bibitem [{\citenamefont {Tongay}\ \emph {et~al.}(2013)\citenamefont {Tongay},
  \citenamefont {Suh}, \citenamefont {Ataca}, \citenamefont {Fan},
  \citenamefont {Luce}, \citenamefont {Kang}, \citenamefont {Liu},
  \citenamefont {Ko}, \citenamefont {Raghunathanan}, \citenamefont {Zhou},
  \citenamefont {Ogletree}, \citenamefont {Li}, \citenamefont {Grossman},\ and\
  \citenamefont {Wu}}]{Tongay2013}%
  \BibitemOpen
  \bibfield  {author} {\bibinfo {author} {\bibfnamefont {Sefaattin}\
  \bibnamefont {Tongay}}, \bibinfo {author} {\bibfnamefont {Joonki}\
  \bibnamefont {Suh}}, \bibinfo {author} {\bibfnamefont {Can}\ \bibnamefont
  {Ataca}}, \bibinfo {author} {\bibfnamefont {Wen}\ \bibnamefont {Fan}},
  \bibinfo {author} {\bibfnamefont {Alexander}\ \bibnamefont {Luce}}, \bibinfo
  {author} {\bibfnamefont {Jeong~Seuk}\ \bibnamefont {Kang}}, \bibinfo {author}
  {\bibfnamefont {Jonathan}\ \bibnamefont {Liu}}, \bibinfo {author}
  {\bibfnamefont {Changhyun}\ \bibnamefont {Ko}}, \bibinfo {author}
  {\bibfnamefont {Rajamani}\ \bibnamefont {Raghunathanan}}, \bibinfo {author}
  {\bibfnamefont {Jian}\ \bibnamefont {Zhou}}, \bibinfo {author} {\bibfnamefont
  {Frank}\ \bibnamefont {Ogletree}}, \bibinfo {author} {\bibfnamefont {Jingbo}\
  \bibnamefont {Li}}, \bibinfo {author} {\bibfnamefont {Jeffrey~C.}\
  \bibnamefont {Grossman}}, \ and\ \bibinfo {author} {\bibfnamefont {Junqiao}\
  \bibnamefont {Wu}},\ }\bibfield  {title} {\enquote {\bibinfo {title} {Defects
  activated photoluminescence in two-dimensional semiconductors: interplay
  between bound, charged and free excitons},}\ }\href
  {https://doi.org/10.1038/srep02657} {\bibfield  {journal} {\bibinfo
  {journal} {Scientific Reports}\ }\textbf {\bibinfo {volume} {3}},\ \bibinfo
  {pages} {2657} (\bibinfo {year} {2013})}\BibitemShut {NoStop}%
\bibitem [{\citenamefont {Chow}\ \emph {et~al.}(2015)\citenamefont {Chow},
  \citenamefont {Jacobs-Gedrim}, \citenamefont {Gao}, \citenamefont {Lu},
  \citenamefont {Yu}, \citenamefont {Terrones},\ and\ \citenamefont
  {Koratkar}}]{Chow2015}%
  \BibitemOpen
  \bibfield  {author} {\bibinfo {author} {\bibfnamefont {Philippe~K.}\
  \bibnamefont {Chow}}, \bibinfo {author} {\bibfnamefont {Robin~B.}\
  \bibnamefont {Jacobs-Gedrim}}, \bibinfo {author} {\bibfnamefont {Jian}\
  \bibnamefont {Gao}}, \bibinfo {author} {\bibfnamefont {Toh-Ming}\
  \bibnamefont {Lu}}, \bibinfo {author} {\bibfnamefont {Bin}\ \bibnamefont
  {Yu}}, \bibinfo {author} {\bibfnamefont {Humberto}\ \bibnamefont {Terrones}},
  \ and\ \bibinfo {author} {\bibfnamefont {Nikhil}\ \bibnamefont {Koratkar}},\
  }\bibfield  {title} {\enquote {\bibinfo {title} {Defect-induced
  photoluminescence in monolayer semiconducting transition metal
  dichalcogenides},}\ }\href {\doibase 10.1021/nn5073495} {\bibfield  {journal}
  {\bibinfo  {journal} {ACS Nano}\ }\textbf {\bibinfo {volume} {9}},\ \bibinfo
  {pages} {1520--1527} (\bibinfo {year} {2015})}\BibitemShut {NoStop}%
\bibitem [{\citenamefont {Hong}\ \emph {et~al.}(2015)\citenamefont {Hong},
  \citenamefont {Hu}, \citenamefont {Probert}, \citenamefont {Li},
  \citenamefont {Lv}, \citenamefont {Yang}, \citenamefont {Gu}, \citenamefont
  {Mao}, \citenamefont {Feng}, \citenamefont {Xie}, \citenamefont {Zhang},
  \citenamefont {Wu}, \citenamefont {Zhang}, \citenamefont {Jin}, \citenamefont
  {Ji}, \citenamefont {Zhang}, \citenamefont {Yuan},\ and\ \citenamefont
  {Zhang}}]{Hong2015}%
  \BibitemOpen
  \bibfield  {author} {\bibinfo {author} {\bibfnamefont {Jinhua}\ \bibnamefont
  {Hong}}, \bibinfo {author} {\bibfnamefont {Zhixin}\ \bibnamefont {Hu}},
  \bibinfo {author} {\bibfnamefont {Matt}\ \bibnamefont {Probert}}, \bibinfo
  {author} {\bibfnamefont {Kun}\ \bibnamefont {Li}}, \bibinfo {author}
  {\bibfnamefont {Danhui}\ \bibnamefont {Lv}}, \bibinfo {author} {\bibfnamefont
  {Xinan}\ \bibnamefont {Yang}}, \bibinfo {author} {\bibfnamefont {Lin}\
  \bibnamefont {Gu}}, \bibinfo {author} {\bibfnamefont {Nannan}\ \bibnamefont
  {Mao}}, \bibinfo {author} {\bibfnamefont {Qingliang}\ \bibnamefont {Feng}},
  \bibinfo {author} {\bibfnamefont {Liming}\ \bibnamefont {Xie}}, \bibinfo
  {author} {\bibfnamefont {Jin}\ \bibnamefont {Zhang}}, \bibinfo {author}
  {\bibfnamefont {Dianzhong}\ \bibnamefont {Wu}}, \bibinfo {author}
  {\bibfnamefont {Zhiyong}\ \bibnamefont {Zhang}}, \bibinfo {author}
  {\bibfnamefont {Chuanhong}\ \bibnamefont {Jin}}, \bibinfo {author}
  {\bibfnamefont {Wei}\ \bibnamefont {Ji}}, \bibinfo {author} {\bibfnamefont
  {Xixiang}\ \bibnamefont {Zhang}}, \bibinfo {author} {\bibfnamefont {Jun}\
  \bibnamefont {Yuan}}, \ and\ \bibinfo {author} {\bibfnamefont
  {Ze}~\bibnamefont {Zhang}},\ }\bibfield  {title} {\enquote {\bibinfo {title}
  {Exploring atomic defects in molybdenum disulphide monolayers},}\ }\href
  {https://doi.org/10.1038/ncomms7293} {\bibfield  {journal} {\bibinfo
  {journal} {Nature Communications}\ }\textbf {\bibinfo {volume} {6}},\
  \bibinfo {pages} {6293} (\bibinfo {year} {2015})}\BibitemShut {NoStop}%
\bibitem [{\citenamefont {Lin}\ \emph {et~al.}(2016)\citenamefont {Lin},
  \citenamefont {Carvalho}, \citenamefont {Kahn}, \citenamefont {Lv},
  \citenamefont {Rao}, \citenamefont {Terrones}, \citenamefont {Pimenta},\ and\
  \citenamefont {Terrones}}]{Lin2016}%
  \BibitemOpen
  \bibfield  {author} {\bibinfo {author} {\bibfnamefont {Zhong}\ \bibnamefont
  {Lin}}, \bibinfo {author} {\bibfnamefont {Bruno~R}\ \bibnamefont {Carvalho}},
  \bibinfo {author} {\bibfnamefont {Ethan}\ \bibnamefont {Kahn}}, \bibinfo
  {author} {\bibfnamefont {Ruitao}\ \bibnamefont {Lv}}, \bibinfo {author}
  {\bibfnamefont {Rahul}\ \bibnamefont {Rao}}, \bibinfo {author} {\bibfnamefont
  {Humberto}\ \bibnamefont {Terrones}}, \bibinfo {author} {\bibfnamefont
  {Marcos~A}\ \bibnamefont {Pimenta}}, \ and\ \bibinfo {author} {\bibfnamefont
  {Mauricio}\ \bibnamefont {Terrones}},\ }\bibfield  {title} {\enquote
  {\bibinfo {title} {Defect engineering of two-dimensional transition metal
  dichalcogenides},}\ }\href {\doibase 10.1088/2053-1583/3/2/022002} {\bibfield
   {journal} {\bibinfo  {journal} {2D Materials}\ }\textbf {\bibinfo {volume}
  {3}},\ \bibinfo {pages} {022002} (\bibinfo {year} {2016})}\BibitemShut
  {NoStop}%
\bibitem [{\citenamefont {Khan}\ \emph {et~al.}(2017)\citenamefont {Khan},
  \citenamefont {Erementchouk}, \citenamefont {Hendrickson},\ and\
  \citenamefont {Leuenberger}}]{Khan2017}%
  \BibitemOpen
  \bibfield  {author} {\bibinfo {author} {\bibfnamefont {M.~A.}\ \bibnamefont
  {Khan}}, \bibinfo {author} {\bibfnamefont {Mikhail}\ \bibnamefont
  {Erementchouk}}, \bibinfo {author} {\bibfnamefont {Joshua}\ \bibnamefont
  {Hendrickson}}, \ and\ \bibinfo {author} {\bibfnamefont {Michael~N.}\
  \bibnamefont {Leuenberger}},\ }\bibfield  {title} {\enquote {\bibinfo {title}
  {Electronic and optical properties of vacancy defects in single-layer
  transition metal dichalcogenides},}\ }\href {\doibase
  10.1103/PhysRevB.95.245435} {\bibfield  {journal} {\bibinfo  {journal} {Phys.
  Rev. B}\ }\textbf {\bibinfo {volume} {95}},\ \bibinfo {pages} {245435}
  (\bibinfo {year} {2017})}\BibitemShut {NoStop}%
\bibitem [{\citenamefont {Klein}\ \emph {et~al.}(2019)\citenamefont {Klein},
  \citenamefont {Lorke}, \citenamefont {Florian}, \citenamefont {Sigger},
  \citenamefont {Sigl}, \citenamefont {Rey}, \citenamefont {Wierzbowski},
  \citenamefont {Cerne}, \citenamefont {Müller}, \citenamefont {Mitterreiter},
  \citenamefont {Zimmermann}, \citenamefont {Taniguchi}, \citenamefont
  {Watanabe}, \citenamefont {Wurstbauer}, \citenamefont {Kaniber},
  \citenamefont {Knap}, \citenamefont {Schmidt}, \citenamefont {Finley},\ and\
  \citenamefont {Holleitner}}]{Klein2019}%
  \BibitemOpen
  \bibfield  {author} {\bibinfo {author} {\bibfnamefont {J.}~\bibnamefont
  {Klein}}, \bibinfo {author} {\bibfnamefont {M.}~\bibnamefont {Lorke}},
  \bibinfo {author} {\bibfnamefont {M.}~\bibnamefont {Florian}}, \bibinfo
  {author} {\bibfnamefont {F.}~\bibnamefont {Sigger}}, \bibinfo {author}
  {\bibfnamefont {L.}~\bibnamefont {Sigl}}, \bibinfo {author} {\bibfnamefont
  {S.}~\bibnamefont {Rey}}, \bibinfo {author} {\bibfnamefont {J.}~\bibnamefont
  {Wierzbowski}}, \bibinfo {author} {\bibfnamefont {J.}~\bibnamefont {Cerne}},
  \bibinfo {author} {\bibfnamefont {K.}~\bibnamefont {Müller}}, \bibinfo
  {author} {\bibfnamefont {E.}~\bibnamefont {Mitterreiter}}, \bibinfo {author}
  {\bibfnamefont {P.}~\bibnamefont {Zimmermann}}, \bibinfo {author}
  {\bibfnamefont {T.}~\bibnamefont {Taniguchi}}, \bibinfo {author}
  {\bibfnamefont {K.}~\bibnamefont {Watanabe}}, \bibinfo {author}
  {\bibfnamefont {U.}~\bibnamefont {Wurstbauer}}, \bibinfo {author}
  {\bibfnamefont {M.}~\bibnamefont {Kaniber}}, \bibinfo {author} {\bibfnamefont
  {M.}~\bibnamefont {Knap}}, \bibinfo {author} {\bibfnamefont {R.}~\bibnamefont
  {Schmidt}}, \bibinfo {author} {\bibfnamefont {J.~J.}\ \bibnamefont {Finley}},
  \ and\ \bibinfo {author} {\bibfnamefont {A.~W.}\ \bibnamefont {Holleitner}},\
  }\bibfield  {title} {\enquote {\bibinfo {title} {Site-selectively generated
  photon emitters in monolayer mos2 via local helium ion irradiation},}\ }\href
  {https://doi.org/10.1038/s41467-019-10632-z} {\bibfield  {journal} {\bibinfo
  {journal} {Nature Communications}\ }\textbf {\bibinfo {volume} {10}},\
  \bibinfo {pages} {2755} (\bibinfo {year} {2019})}\BibitemShut {NoStop}%
\bibitem [{\citenamefont {Mitterreiter}\ \emph {et~al.}(2021)\citenamefont
  {Mitterreiter}, \citenamefont {Schuler}, \citenamefont {Micevic},
  \citenamefont {Hernangómez-Pérez}, \citenamefont {Barthelmi}, \citenamefont
  {Cochrane}, \citenamefont {Kiemle}, \citenamefont {Sigger}, \citenamefont
  {Klein}, \citenamefont {Wong}, \citenamefont {Barnard}, \citenamefont
  {Watanabe}, \citenamefont {Taniguchi}, \citenamefont {Lorke}, \citenamefont
  {Jahnke}, \citenamefont {Finley}, \citenamefont {Schwartzberg}, \citenamefont
  {Qiu}, \citenamefont {Refaely-Abramson}, \citenamefont {Holleitner},
  \citenamefont {Weber-Bargioni},\ and\ \citenamefont
  {Kastl}}]{Mitterreiter2021}%
  \BibitemOpen
  \bibfield  {author} {\bibinfo {author} {\bibfnamefont {Elmar}\ \bibnamefont
  {Mitterreiter}}, \bibinfo {author} {\bibfnamefont {Bruno}\ \bibnamefont
  {Schuler}}, \bibinfo {author} {\bibfnamefont {Ana}\ \bibnamefont {Micevic}},
  \bibinfo {author} {\bibfnamefont {Daniel}\ \bibnamefont
  {Hernangómez-Pérez}}, \bibinfo {author} {\bibfnamefont {Katja}\
  \bibnamefont {Barthelmi}}, \bibinfo {author} {\bibfnamefont {Katherine~A.}\
  \bibnamefont {Cochrane}}, \bibinfo {author} {\bibfnamefont {Jonas}\
  \bibnamefont {Kiemle}}, \bibinfo {author} {\bibfnamefont {Florian}\
  \bibnamefont {Sigger}}, \bibinfo {author} {\bibfnamefont {Julian}\
  \bibnamefont {Klein}}, \bibinfo {author} {\bibfnamefont {Edward}\
  \bibnamefont {Wong}}, \bibinfo {author} {\bibfnamefont {Edward~S.}\
  \bibnamefont {Barnard}}, \bibinfo {author} {\bibfnamefont {Kenji}\
  \bibnamefont {Watanabe}}, \bibinfo {author} {\bibfnamefont {Takashi}\
  \bibnamefont {Taniguchi}}, \bibinfo {author} {\bibfnamefont {Michael}\
  \bibnamefont {Lorke}}, \bibinfo {author} {\bibfnamefont {Frank}\ \bibnamefont
  {Jahnke}}, \bibinfo {author} {\bibfnamefont {Johnathan~J.}\ \bibnamefont
  {Finley}}, \bibinfo {author} {\bibfnamefont {Adam~M.}\ \bibnamefont
  {Schwartzberg}}, \bibinfo {author} {\bibfnamefont {Diana~Y.}\ \bibnamefont
  {Qiu}}, \bibinfo {author} {\bibfnamefont {Sivan}\ \bibnamefont
  {Refaely-Abramson}}, \bibinfo {author} {\bibfnamefont {Alexander~W.}\
  \bibnamefont {Holleitner}}, \bibinfo {author} {\bibfnamefont {Alexander}\
  \bibnamefont {Weber-Bargioni}}, \ and\ \bibinfo {author} {\bibfnamefont
  {Christoph}\ \bibnamefont {Kastl}},\ }\bibfield  {title} {\enquote {\bibinfo
  {title} {The role of chalcogen vacancies for atomic defect emission in
  mos2},}\ }\href {https://doi.org/10.1038/s41467-021-24102-y} {\bibfield
  {journal} {\bibinfo  {journal} {Nature Communications}\ }\textbf {\bibinfo
  {volume} {12}},\ \bibinfo {pages} {3822} (\bibinfo {year}
  {2021})}\BibitemShut {NoStop}%
\bibitem [{\citenamefont {Rosenberger}\ \emph {et~al.}(2018)\citenamefont
  {Rosenberger}, \citenamefont {Chuang}, \citenamefont {McCreary},
  \citenamefont {Li},\ and\ \citenamefont {Jonker}}]{Rosenberger2018}%
  \BibitemOpen
  \bibfield  {author} {\bibinfo {author} {\bibfnamefont {Matthew~R.}\
  \bibnamefont {Rosenberger}}, \bibinfo {author} {\bibfnamefont {Hsun-Jen}\
  \bibnamefont {Chuang}}, \bibinfo {author} {\bibfnamefont {Kathleen~M.}\
  \bibnamefont {McCreary}}, \bibinfo {author} {\bibfnamefont {Connie~H.}\
  \bibnamefont {Li}}, \ and\ \bibinfo {author} {\bibfnamefont {Berend~T.}\
  \bibnamefont {Jonker}},\ }\bibfield  {title} {\enquote {\bibinfo {title}
  {Electrical characterization of discrete defects and impact of defect density
  on photoluminescence in monolayer ws2},}\ }\href {\doibase
  10.1021/acsnano.7b08566} {\bibfield  {journal} {\bibinfo  {journal} {ACS
  Nano}\ }\textbf {\bibinfo {volume} {12}},\ \bibinfo {pages} {1793--1800}
  (\bibinfo {year} {2018})}\BibitemShut {NoStop}%
\bibitem [{\citenamefont {Greben}\ \emph {et~al.}(2020)\citenamefont {Greben},
  \citenamefont {Arora}, \citenamefont {Harats},\ and\ \citenamefont
  {Bolotin}}]{Greben2020}%
  \BibitemOpen
  \bibfield  {author} {\bibinfo {author} {\bibfnamefont {Kyrylo}\ \bibnamefont
  {Greben}}, \bibinfo {author} {\bibfnamefont {Sonakshi}\ \bibnamefont
  {Arora}}, \bibinfo {author} {\bibfnamefont {Moshe~G.}\ \bibnamefont
  {Harats}}, \ and\ \bibinfo {author} {\bibfnamefont {Kirill~I.}\ \bibnamefont
  {Bolotin}},\ }\bibfield  {title} {\enquote {\bibinfo {title} {Intrinsic and
  extrinsic defect-related excitons in tmdcs},}\ }\href {\doibase
  10.1021/acs.nanolett.9b05323} {\bibfield  {journal} {\bibinfo  {journal}
  {Nano Lett.}\ }\textbf {\bibinfo {volume} {20}},\ \bibinfo {pages}
  {2544--2550} (\bibinfo {year} {2020})}\BibitemShut {NoStop}%
\bibitem [{\citenamefont {Wu}\ \emph {et~al.}(2022)\citenamefont {Wu},
  \citenamefont {Zhong}, \citenamefont {Guo}, \citenamefont {Tang},
  \citenamefont {Yang}, \citenamefont {Qian}, \citenamefont {Yuan},
  \citenamefont {Zhang},\ and\ \citenamefont {Xu}}]{Wu2022}%
  \BibitemOpen
  \bibfield  {author} {\bibinfo {author} {\bibfnamefont {Ke}~\bibnamefont
  {Wu}}, \bibinfo {author} {\bibfnamefont {Hongxia}\ \bibnamefont {Zhong}},
  \bibinfo {author} {\bibfnamefont {Quanbing}\ \bibnamefont {Guo}}, \bibinfo
  {author} {\bibfnamefont {Jibo}\ \bibnamefont {Tang}}, \bibinfo {author}
  {\bibfnamefont {Zhenyu}\ \bibnamefont {Yang}}, \bibinfo {author}
  {\bibfnamefont {Lihua}\ \bibnamefont {Qian}}, \bibinfo {author}
  {\bibfnamefont {Shengjun}\ \bibnamefont {Yuan}}, \bibinfo {author}
  {\bibfnamefont {Shunping}\ \bibnamefont {Zhang}}, \ and\ \bibinfo {author}
  {\bibfnamefont {Hongxing}\ \bibnamefont {Xu}},\ }\bibfield  {title} {\enquote
  {\bibinfo {title} {Revealing the competition between defect-trapped exciton
  and band-edge exciton photoluminescence in monolayer hexagonal ws2},}\ }\href
  {\doibase https://doi.org/10.1002/adom.202101971} {\bibfield  {journal}
  {\bibinfo  {journal} {Advanced Optical Materials}\ }\textbf {\bibinfo
  {volume} {10}},\ \bibinfo {pages} {2101971} (\bibinfo {year} {2022})},\
  \Eprint
  {http://arxiv.org/abs/https://advanced.onlinelibrary.wiley.com/doi/pdf/10.1002/adom.202101971}
  {https://advanced.onlinelibrary.wiley.com/doi/pdf/10.1002/adom.202101971}
  \BibitemShut {NoStop}%
\bibitem [{\citenamefont {Koperski}\ \emph {et~al.}(2015)\citenamefont
  {Koperski}, \citenamefont {Nogajewski}, \citenamefont {Arora}, \citenamefont
  {Cherkez}, \citenamefont {Mallet}, \citenamefont {Veuillen}, \citenamefont
  {Marcus}, \citenamefont {Kossacki},\ and\ \citenamefont
  {Potemski}}]{Koperski2015}%
  \BibitemOpen
  \bibfield  {author} {\bibinfo {author} {\bibfnamefont {M.}~\bibnamefont
  {Koperski}}, \bibinfo {author} {\bibfnamefont {K.}~\bibnamefont
  {Nogajewski}}, \bibinfo {author} {\bibfnamefont {A.}~\bibnamefont {Arora}},
  \bibinfo {author} {\bibfnamefont {V.}~\bibnamefont {Cherkez}}, \bibinfo
  {author} {\bibfnamefont {P.}~\bibnamefont {Mallet}}, \bibinfo {author}
  {\bibfnamefont {J.-Y.}\ \bibnamefont {Veuillen}}, \bibinfo {author}
  {\bibfnamefont {J.}~\bibnamefont {Marcus}}, \bibinfo {author} {\bibfnamefont
  {P.}~\bibnamefont {Kossacki}}, \ and\ \bibinfo {author} {\bibfnamefont
  {M.}~\bibnamefont {Potemski}},\ }\bibfield  {title} {\enquote {\bibinfo
  {title} {Single photon emitters in exfoliated wse2 structures},}\ }\href
  {https://doi.org/10.1038/nnano.2015.67} {\bibfield  {journal} {\bibinfo
  {journal} {Nature Nanotechnology}\ }\textbf {\bibinfo {volume} {10}},\
  \bibinfo {pages} {503--506} (\bibinfo {year} {2015})}\BibitemShut {NoStop}%
\bibitem [{\citenamefont {He}\ \emph {et~al.}(2015)\citenamefont {He},
  \citenamefont {Clark}, \citenamefont {Schaibley}, \citenamefont {He},
  \citenamefont {Chen}, \citenamefont {Wei}, \citenamefont {Ding},
  \citenamefont {Zhang}, \citenamefont {Yao}, \citenamefont {Xu}, \citenamefont
  {Lu},\ and\ \citenamefont {Pan}}]{He2015}%
  \BibitemOpen
  \bibfield  {author} {\bibinfo {author} {\bibfnamefont {Yu-Ming}\ \bibnamefont
  {He}}, \bibinfo {author} {\bibfnamefont {Genevieve}\ \bibnamefont {Clark}},
  \bibinfo {author} {\bibfnamefont {John~R.}\ \bibnamefont {Schaibley}},
  \bibinfo {author} {\bibfnamefont {Yu}~\bibnamefont {He}}, \bibinfo {author}
  {\bibfnamefont {Ming-Cheng}\ \bibnamefont {Chen}}, \bibinfo {author}
  {\bibfnamefont {Yu-Jia}\ \bibnamefont {Wei}}, \bibinfo {author}
  {\bibfnamefont {Xing}\ \bibnamefont {Ding}}, \bibinfo {author} {\bibfnamefont
  {Qiang}\ \bibnamefont {Zhang}}, \bibinfo {author} {\bibfnamefont {Wang}\
  \bibnamefont {Yao}}, \bibinfo {author} {\bibfnamefont {Xiaodong}\
  \bibnamefont {Xu}}, \bibinfo {author} {\bibfnamefont {Chao-Yang}\
  \bibnamefont {Lu}}, \ and\ \bibinfo {author} {\bibfnamefont {Jian-Wei}\
  \bibnamefont {Pan}},\ }\bibfield  {title} {\enquote {\bibinfo {title} {Single
  quantum emitters in monolayer semiconductors},}\ }\href
  {https://doi.org/10.1038/nnano.2015.75} {\bibfield  {journal} {\bibinfo
  {journal} {Nature Nanotechnology}\ }\textbf {\bibinfo {volume} {10}},\
  \bibinfo {pages} {497--502} (\bibinfo {year} {2015})}\BibitemShut {NoStop}%
\bibitem [{\citenamefont {Srivastava}\ \emph {et~al.}(2015)\citenamefont
  {Srivastava}, \citenamefont {Sidler}, \citenamefont {Allain}, \citenamefont
  {Lembke}, \citenamefont {Kis},\ and\ \citenamefont
  {Imamoğlu}}]{Srivastava2015}%
  \BibitemOpen
  \bibfield  {author} {\bibinfo {author} {\bibfnamefont {Ajit}\ \bibnamefont
  {Srivastava}}, \bibinfo {author} {\bibfnamefont {Meinrad}\ \bibnamefont
  {Sidler}}, \bibinfo {author} {\bibfnamefont {Adrien~V.}\ \bibnamefont
  {Allain}}, \bibinfo {author} {\bibfnamefont {Dominik~S.}\ \bibnamefont
  {Lembke}}, \bibinfo {author} {\bibfnamefont {Andras}\ \bibnamefont {Kis}}, \
  and\ \bibinfo {author} {\bibfnamefont {A.}~\bibnamefont {Imamoğlu}},\
  }\bibfield  {title} {\enquote {\bibinfo {title} {Optically active quantum
  dots in monolayer wse2},}\ }\href {https://doi.org/10.1038/nnano.2015.60}
  {\bibfield  {journal} {\bibinfo  {journal} {Nature Nanotechnology}\ }\textbf
  {\bibinfo {volume} {10}},\ \bibinfo {pages} {491--496} (\bibinfo {year}
  {2015})}\BibitemShut {NoStop}%
\bibitem [{\citenamefont {Chakraborty}\ \emph {et~al.}(2015)\citenamefont
  {Chakraborty}, \citenamefont {Kinnischtzke}, \citenamefont {Goodfellow},
  \citenamefont {Beams},\ and\ \citenamefont {Vamivakas}}]{Chakraborty2015}%
  \BibitemOpen
  \bibfield  {author} {\bibinfo {author} {\bibfnamefont {Chitraleema}\
  \bibnamefont {Chakraborty}}, \bibinfo {author} {\bibfnamefont {Laura}\
  \bibnamefont {Kinnischtzke}}, \bibinfo {author} {\bibfnamefont {Kenneth~M.}\
  \bibnamefont {Goodfellow}}, \bibinfo {author} {\bibfnamefont {Ryan}\
  \bibnamefont {Beams}}, \ and\ \bibinfo {author} {\bibfnamefont {A.~Nick}\
  \bibnamefont {Vamivakas}},\ }\bibfield  {title} {\enquote {\bibinfo {title}
  {Voltage-controlled quantum light from an atomically thin semiconductor},}\
  }\href {https://doi.org/10.1038/nnano.2015.79} {\bibfield  {journal}
  {\bibinfo  {journal} {Nature Nanotechnology}\ }\textbf {\bibinfo {volume}
  {10}},\ \bibinfo {pages} {507--511} (\bibinfo {year} {2015})}\BibitemShut
  {NoStop}%
\bibitem [{\citenamefont {Chakraborty}\ \emph {et~al.}(2019)\citenamefont
  {Chakraborty}, \citenamefont {Vamivakas},\ and\ \citenamefont
  {Englund}}]{ChakrabortyVamivakasEnglund2019}%
  \BibitemOpen
  \bibfield  {author} {\bibinfo {author} {\bibfnamefont {Chitraleema}\
  \bibnamefont {Chakraborty}}, \bibinfo {author} {\bibfnamefont {Nick}\
  \bibnamefont {Vamivakas}}, \ and\ \bibinfo {author} {\bibfnamefont {Dirk}\
  \bibnamefont {Englund}},\ }\bibfield  {title} {\enquote {\bibinfo {title}
  {Advances in quantum light emission from 2d materials},}\ }\href {\doibase
  doi:10.1515/nanoph-2019-0140} {\bibfield  {journal} {\bibinfo  {journal}
  {Nanophotonics}\ }\textbf {\bibinfo {volume} {8}},\ \bibinfo {pages}
  {2017--2032} (\bibinfo {year} {2019})}\BibitemShut {NoStop}%
\bibitem [{\citenamefont {Barthelmi}\ \emph {et~al.}(2020)\citenamefont
  {Barthelmi}, \citenamefont {Klein}, \citenamefont {Hötger}, \citenamefont
  {Sigl}, \citenamefont {Sigger}, \citenamefont {Mitterreiter}, \citenamefont
  {Rey}, \citenamefont {Gyger}, \citenamefont {Lorke}, \citenamefont {Florian},
  \citenamefont {Jahnke}, \citenamefont {Taniguchi}, \citenamefont {Watanabe},
  \citenamefont {Zwiller}, \citenamefont {Jöns}, \citenamefont {Wurstbauer},
  \citenamefont {Kastl}, \citenamefont {Weber-Bargioni}, \citenamefont
  {Finley}, \citenamefont {Müller},\ and\ \citenamefont
  {Holleitner}}]{Barthelmi2020}%
  \BibitemOpen
  \bibfield  {author} {\bibinfo {author} {\bibfnamefont {K.}~\bibnamefont
  {Barthelmi}}, \bibinfo {author} {\bibfnamefont {J.}~\bibnamefont {Klein}},
  \bibinfo {author} {\bibfnamefont {A.}~\bibnamefont {Hötger}}, \bibinfo
  {author} {\bibfnamefont {L.}~\bibnamefont {Sigl}}, \bibinfo {author}
  {\bibfnamefont {F.}~\bibnamefont {Sigger}}, \bibinfo {author} {\bibfnamefont
  {E.}~\bibnamefont {Mitterreiter}}, \bibinfo {author} {\bibfnamefont
  {S.}~\bibnamefont {Rey}}, \bibinfo {author} {\bibfnamefont {S.}~\bibnamefont
  {Gyger}}, \bibinfo {author} {\bibfnamefont {M.}~\bibnamefont {Lorke}},
  \bibinfo {author} {\bibfnamefont {M.}~\bibnamefont {Florian}}, \bibinfo
  {author} {\bibfnamefont {F.}~\bibnamefont {Jahnke}}, \bibinfo {author}
  {\bibfnamefont {T.}~\bibnamefont {Taniguchi}}, \bibinfo {author}
  {\bibfnamefont {K.}~\bibnamefont {Watanabe}}, \bibinfo {author}
  {\bibfnamefont {V.}~\bibnamefont {Zwiller}}, \bibinfo {author} {\bibfnamefont
  {K.~D.}\ \bibnamefont {Jöns}}, \bibinfo {author} {\bibfnamefont
  {U.}~\bibnamefont {Wurstbauer}}, \bibinfo {author} {\bibfnamefont
  {C.}~\bibnamefont {Kastl}}, \bibinfo {author} {\bibfnamefont
  {A.}~\bibnamefont {Weber-Bargioni}}, \bibinfo {author} {\bibfnamefont
  {J.~J.}\ \bibnamefont {Finley}}, \bibinfo {author} {\bibfnamefont
  {K.}~\bibnamefont {Müller}}, \ and\ \bibinfo {author} {\bibfnamefont
  {A.~W.}\ \bibnamefont {Holleitner}},\ }\bibfield  {title} {\enquote {\bibinfo
  {title} {Atomistic defects as single-photon emitters in atomically thin
  mos2},}\ }\href {https://doi.org/10.1063/5.0018557} {\bibfield  {journal}
  {\bibinfo  {journal} {Appl. Phys. Lett.}\ }\textbf {\bibinfo {volume}
  {117}},\ \bibinfo {pages} {070501} (\bibinfo {year} {2020})}\BibitemShut
  {NoStop}%
\bibitem [{\citenamefont {Attaccalite}\ \emph {et~al.}(2011)\citenamefont
  {Attaccalite}, \citenamefont {Bockstedte}, \citenamefont {Marini},
  \citenamefont {Rubio},\ and\ \citenamefont {Wirtz}}]{Attaccalite2011}%
  \BibitemOpen
  \bibfield  {author} {\bibinfo {author} {\bibfnamefont {C.}~\bibnamefont
  {Attaccalite}}, \bibinfo {author} {\bibfnamefont {M.}~\bibnamefont
  {Bockstedte}}, \bibinfo {author} {\bibfnamefont {A.}~\bibnamefont {Marini}},
  \bibinfo {author} {\bibfnamefont {A.}~\bibnamefont {Rubio}}, \ and\ \bibinfo
  {author} {\bibfnamefont {L.}~\bibnamefont {Wirtz}},\ }\bibfield  {title}
  {\enquote {\bibinfo {title} {Coupling of excitons and defect states in
  boron-nitride nanostructures},}\ }\href {\doibase 10.1103/PhysRevB.83.144115}
  {\bibfield  {journal} {\bibinfo  {journal} {Phys. Rev. B}\ }\textbf {\bibinfo
  {volume} {83}},\ \bibinfo {pages} {144115} (\bibinfo {year}
  {2011})}\BibitemShut {NoStop}%
\bibitem [{\citenamefont {Refaely-Abramson}\ \emph {et~al.}(2018)\citenamefont
  {Refaely-Abramson}, \citenamefont {Qiu}, \citenamefont {Louie},\ and\
  \citenamefont {Neaton}}]{Sivan2018}%
  \BibitemOpen
  \bibfield  {author} {\bibinfo {author} {\bibfnamefont {Sivan}\ \bibnamefont
  {Refaely-Abramson}}, \bibinfo {author} {\bibfnamefont {Diana~Y.}\
  \bibnamefont {Qiu}}, \bibinfo {author} {\bibfnamefont {Steven~G.}\
  \bibnamefont {Louie}}, \ and\ \bibinfo {author} {\bibfnamefont {Jeffrey~B.}\
  \bibnamefont {Neaton}},\ }\bibfield  {title} {\enquote {\bibinfo {title}
  {Defect-induced modification of low-lying excitons and valley selectivity in
  monolayer transition metal dichalcogenides},}\ }\href {\doibase
  10.1103/PhysRevLett.121.167402} {\bibfield  {journal} {\bibinfo  {journal}
  {Phys. Rev. Lett.}\ }\textbf {\bibinfo {volume} {121}},\ \bibinfo {pages}
  {167402} (\bibinfo {year} {2018})}\BibitemShut {NoStop}%
\bibitem [{\citenamefont {Zhou}\ \emph {et~al.}(2021)\citenamefont {Zhou},
  \citenamefont {Park}, \citenamefont {Lu}, \citenamefont {Maliyov},
  \citenamefont {Tong},\ and\ \citenamefont {Bernardi}}]{Perturbo}%
  \BibitemOpen
  \bibfield  {author} {\bibinfo {author} {\bibfnamefont {Jin-Jian}\
  \bibnamefont {Zhou}}, \bibinfo {author} {\bibfnamefont {Jinsoo}\ \bibnamefont
  {Park}}, \bibinfo {author} {\bibfnamefont {I-Te}\ \bibnamefont {Lu}},
  \bibinfo {author} {\bibfnamefont {Ivan}\ \bibnamefont {Maliyov}}, \bibinfo
  {author} {\bibfnamefont {Xiao}\ \bibnamefont {Tong}}, \ and\ \bibinfo
  {author} {\bibfnamefont {Marco}\ \bibnamefont {Bernardi}},\ }\bibfield
  {title} {\enquote {\bibinfo {title} {Perturbo: A software package for ab
  initio electron–phonon interactions, charge transport and ultrafast
  dynamics},}\ }\href {\doibase https://doi.org/10.1016/j.cpc.2021.107970}
  {\bibfield  {journal} {\bibinfo  {journal} {Computer Physics Communications}\
  }\textbf {\bibinfo {volume} {264}},\ \bibinfo {pages} {107970} (\bibinfo
  {year} {2021})}\BibitemShut {NoStop}%
\bibitem [{\citenamefont {Lu}\ \emph {et~al.}(2019)\citenamefont {Lu},
  \citenamefont {Zhou},\ and\ \citenamefont {Bernardi}}]{Lu2019}%
  \BibitemOpen
  \bibfield  {author} {\bibinfo {author} {\bibfnamefont {I-Te}\ \bibnamefont
  {Lu}}, \bibinfo {author} {\bibfnamefont {Jin-Jian}\ \bibnamefont {Zhou}}, \
  and\ \bibinfo {author} {\bibfnamefont {Marco}\ \bibnamefont {Bernardi}},\
  }\bibfield  {title} {\enquote {\bibinfo {title} {Efficient ab initio
  calculations of electron-defect scattering and defect-limited carrier
  mobility},}\ }\href {\doibase 10.1103/PhysRevMaterials.3.033804} {\bibfield
  {journal} {\bibinfo  {journal} {Phys. Rev. Mater.}\ }\textbf {\bibinfo
  {volume} {3}},\ \bibinfo {pages} {033804} (\bibinfo {year}
  {2019})}\BibitemShut {NoStop}%
\bibitem [{\citenamefont {Lu}\ \emph {et~al.}(2020)\citenamefont {Lu},
  \citenamefont {Park}, \citenamefont {Zhou},\ and\ \citenamefont
  {Bernardi}}]{Lu2020}%
  \BibitemOpen
  \bibfield  {author} {\bibinfo {author} {\bibfnamefont {I.-Te}\ \bibnamefont
  {Lu}}, \bibinfo {author} {\bibfnamefont {Jinsoo}\ \bibnamefont {Park}},
  \bibinfo {author} {\bibfnamefont {Jin-Jian}\ \bibnamefont {Zhou}}, \ and\
  \bibinfo {author} {\bibfnamefont {Marco}\ \bibnamefont {Bernardi}},\
  }\bibfield  {title} {\enquote {\bibinfo {title} {Ab initio electron-defect
  interactions using wannier functions},}\ }\href
  {https://doi.org/10.1038/s41524-020-0284-y} {\bibfield  {journal} {\bibinfo
  {journal} {npj Computational Materials}\ }\textbf {\bibinfo {volume} {6}},\
  \bibinfo {pages} {17} (\bibinfo {year} {2020})}\BibitemShut {NoStop}%
\bibitem [{\citenamefont {Kaasbjerg}(2020)}]{Kristen2020}%
  \BibitemOpen
  \bibfield  {author} {\bibinfo {author} {\bibfnamefont {Kristen}\ \bibnamefont
  {Kaasbjerg}},\ }\bibfield  {title} {\enquote {\bibinfo {title} {Atomistic
  $t$-matrix theory of disordered two-dimensional materials: Bound states,
  spectral properties, quasiparticle scattering, and transport},}\ }\href
  {\doibase 10.1103/PhysRevB.101.045433} {\bibfield  {journal} {\bibinfo
  {journal} {Phys. Rev. B}\ }\textbf {\bibinfo {volume} {101}},\ \bibinfo
  {pages} {045433} (\bibinfo {year} {2020})}\BibitemShut {NoStop}%
\bibitem [{\citenamefont {Xu}\ \emph {et~al.}(2021)\citenamefont {Xu},
  \citenamefont {Habib}, \citenamefont {Sundararaman},\ and\ \citenamefont
  {Ping}}]{Xu2021}%
  \BibitemOpen
  \bibfield  {author} {\bibinfo {author} {\bibfnamefont {Junqing}\ \bibnamefont
  {Xu}}, \bibinfo {author} {\bibfnamefont {Adela}\ \bibnamefont {Habib}},
  \bibinfo {author} {\bibfnamefont {Ravishankar}\ \bibnamefont {Sundararaman}},
  \ and\ \bibinfo {author} {\bibfnamefont {Yuan}\ \bibnamefont {Ping}},\
  }\bibfield  {title} {\enquote {\bibinfo {title} {Ab initio ultrafast spin
  dynamics in solids},}\ }\href {\doibase 10.1103/PhysRevB.104.184418}
  {\bibfield  {journal} {\bibinfo  {journal} {Phys. Rev. B}\ }\textbf {\bibinfo
  {volume} {104}},\ \bibinfo {pages} {184418} (\bibinfo {year}
  {2021})}\BibitemShut {NoStop}%
\bibitem [{\citenamefont {Deslippe}\ \emph {et~al.}(2012)\citenamefont
  {Deslippe}, \citenamefont {Samsonidze}, \citenamefont {Strubbe},
  \citenamefont {Jain}, \citenamefont {Cohen},\ and\ \citenamefont
  {Louie}}]{Deslippe2012}%
  \BibitemOpen
  \bibfield  {author} {\bibinfo {author} {\bibfnamefont {Jack}\ \bibnamefont
  {Deslippe}}, \bibinfo {author} {\bibfnamefont {Georgy}\ \bibnamefont
  {Samsonidze}}, \bibinfo {author} {\bibfnamefont {David~A.}\ \bibnamefont
  {Strubbe}}, \bibinfo {author} {\bibfnamefont {Manish}\ \bibnamefont {Jain}},
  \bibinfo {author} {\bibfnamefont {Marvin~L.}\ \bibnamefont {Cohen}}, \ and\
  \bibinfo {author} {\bibfnamefont {Steven~G.}\ \bibnamefont {Louie}},\
  }\bibfield  {title} {\enquote {\bibinfo {title} {Berkeleygw: A massively
  parallel computer package for the calculation of the quasiparticle and
  optical properties of materials and nanostructures},}\ }\href
  {http://www.sciencedirect.com/science/article/pii/S0010465511003912}
  {\bibfield  {journal} {\bibinfo  {journal} {Computer Physics Communications}\
  }\textbf {\bibinfo {volume} {183}},\ \bibinfo {pages} {1269--1289} (\bibinfo
  {year} {2012})}\BibitemShut {NoStop}%
\bibitem [{\citenamefont {Hybertsen}\ and\ \citenamefont
  {Louie}(1986)}]{Hybertsen1986}%
  \BibitemOpen
  \bibfield  {author} {\bibinfo {author} {\bibfnamefont {Mark~S.}\ \bibnamefont
  {Hybertsen}}\ and\ \bibinfo {author} {\bibfnamefont {Steven~G.}\ \bibnamefont
  {Louie}},\ }\bibfield  {title} {\enquote {\bibinfo {title} {Electron
  correlation in semiconductors and insulators: Band gaps and quasiparticle
  energies},}\ }\href {\doibase 10.1103/PhysRevB.34.5390} {\bibfield  {journal}
  {\bibinfo  {journal} {Phys. Rev. B}\ }\textbf {\bibinfo {volume} {34}},\
  \bibinfo {pages} {5390--5413} (\bibinfo {year} {1986})}\BibitemShut {NoStop}%
\bibitem [{\citenamefont {Rohlfing}\ and\ \citenamefont
  {Louie}(2000)}]{Rohlfing2000}%
  \BibitemOpen
  \bibfield  {author} {\bibinfo {author} {\bibfnamefont {Michael}\ \bibnamefont
  {Rohlfing}}\ and\ \bibinfo {author} {\bibfnamefont {Steven~G.}\ \bibnamefont
  {Louie}},\ }\bibfield  {title} {\enquote {\bibinfo {title} {Electron-hole
  excitations and optical spectra from first principles},}\ }\href {\doibase
  10.1103/PhysRevB.62.4927} {\bibfield  {journal} {\bibinfo  {journal} {Phys.
  Rev. B}\ }\textbf {\bibinfo {volume} {62}},\ \bibinfo {pages} {4927--4944}
  (\bibinfo {year} {2000})}\BibitemShut {NoStop}%
\bibitem [{SM()}]{SM}%
  \BibitemOpen
  \href@noop {} {}\bibinfo {note} {See Supplemental Material at [URL will be
  inserted by publisher].}\BibitemShut {Stop}%
\bibitem [{\citenamefont {Cao}\ \emph {et~al.}(2012)\citenamefont {Cao},
  \citenamefont {Wang}, \citenamefont {Han}, \citenamefont {Ye}, \citenamefont
  {Zhu}, \citenamefont {Shi}, \citenamefont {Niu}, \citenamefont {Tan},
  \citenamefont {Wang}, \citenamefont {Liu},\ and\ \citenamefont
  {Feng}}]{Cao2012}%
  \BibitemOpen
  \bibfield  {author} {\bibinfo {author} {\bibfnamefont {Ting}\ \bibnamefont
  {Cao}}, \bibinfo {author} {\bibfnamefont {Gang}\ \bibnamefont {Wang}},
  \bibinfo {author} {\bibfnamefont {Wenpeng}\ \bibnamefont {Han}}, \bibinfo
  {author} {\bibfnamefont {Huiqi}\ \bibnamefont {Ye}}, \bibinfo {author}
  {\bibfnamefont {Chuanrui}\ \bibnamefont {Zhu}}, \bibinfo {author}
  {\bibfnamefont {Junren}\ \bibnamefont {Shi}}, \bibinfo {author}
  {\bibfnamefont {Qian}\ \bibnamefont {Niu}}, \bibinfo {author} {\bibfnamefont
  {Pingheng}\ \bibnamefont {Tan}}, \bibinfo {author} {\bibfnamefont {Enge}\
  \bibnamefont {Wang}}, \bibinfo {author} {\bibfnamefont {Baoli}\ \bibnamefont
  {Liu}}, \ and\ \bibinfo {author} {\bibfnamefont {Ji}~\bibnamefont {Feng}},\
  }\bibfield  {title} {\enquote {\bibinfo {title} {Valley-selective circular
  dichroism of monolayer molybdenum disulphide},}\ }\href
  {https://doi.org/10.1038/ncomms1882} {\bibfield  {journal} {\bibinfo
  {journal} {Nature Communications}\ }\textbf {\bibinfo {volume} {3}},\
  \bibinfo {pages} {887} (\bibinfo {year} {2012})}\BibitemShut {NoStop}%
\bibitem [{\citenamefont {Chan}\ \emph {et~al.}(2023)\citenamefont {Chan},
  \citenamefont {Haber}, \citenamefont {Naik}, \citenamefont {Neaton},
  \citenamefont {Qiu}, \citenamefont {da~Jornada},\ and\ \citenamefont
  {Louie}}]{Chan2023}%
  \BibitemOpen
  \bibfield  {author} {\bibinfo {author} {\bibfnamefont {Yang-hao}\
  \bibnamefont {Chan}}, \bibinfo {author} {\bibfnamefont {Jonah~B.}\
  \bibnamefont {Haber}}, \bibinfo {author} {\bibfnamefont {Mit~H.}\
  \bibnamefont {Naik}}, \bibinfo {author} {\bibfnamefont {Jeffrey~B.}\
  \bibnamefont {Neaton}}, \bibinfo {author} {\bibfnamefont {Diana~Y.}\
  \bibnamefont {Qiu}}, \bibinfo {author} {\bibfnamefont {Felipe~H.}\
  \bibnamefont {da~Jornada}}, \ and\ \bibinfo {author} {\bibfnamefont
  {Steven~G.}\ \bibnamefont {Louie}},\ }\bibfield  {title} {\enquote {\bibinfo
  {title} {Exciton lifetime and optical line width profile via exciton-phonon
  interactions: Theory and first-principles calculations for monolayer mos2},}\
  }\href {\doibase 10.1021/acs.nanolett.3c00732} {\bibfield  {journal}
  {\bibinfo  {journal} {Nano Lett.}\ }\textbf {\bibinfo {volume} {23}},\
  \bibinfo {pages} {3971--3977} (\bibinfo {year} {2023})}\BibitemShut {NoStop}%
\bibitem [{\citenamefont {Bruus}\ and\ \citenamefont
  {Flensberg}(2004)}]{Bruus2004}%
  \BibitemOpen
  \bibfield  {author} {\bibinfo {author} {\bibfnamefont {Henrik~Bruus}\
  \bibnamefont {Bruus}}\ and\ \bibinfo {author} {\bibfnamefont {Karsten}\
  \bibnamefont {Flensberg}},\ }\href@noop {} {\emph {\bibinfo {title}
  {Many-Body Quantum Theory in Condensed Matter Physics: An Introduction}}}\
  (\bibinfo  {publisher} {Oxford University Press},\ \bibinfo {year}
  {2004})\BibitemShut {NoStop}%
\bibitem [{\citenamefont {Hannewald}\ \emph {et~al.}(2000)\citenamefont
  {Hannewald}, \citenamefont {Glutsch},\ and\ \citenamefont
  {Bechstedt}}]{Hannewald2000}%
  \BibitemOpen
  \bibfield  {author} {\bibinfo {author} {\bibfnamefont {K.}~\bibnamefont
  {Hannewald}}, \bibinfo {author} {\bibfnamefont {S.}~\bibnamefont {Glutsch}},
  \ and\ \bibinfo {author} {\bibfnamefont {F.}~\bibnamefont {Bechstedt}},\
  }\bibfield  {title} {\enquote {\bibinfo {title} {Theory of photoluminescence
  in semiconductors},}\ }\href {\doibase 10.1103/PhysRevB.62.4519} {\bibfield
  {journal} {\bibinfo  {journal} {Phys. Rev. B}\ }\textbf {\bibinfo {volume}
  {62}},\ \bibinfo {pages} {4519--4525} (\bibinfo {year} {2000})}\BibitemShut
  {NoStop}%
\bibitem [{\citenamefont {Marini}\ \emph {et~al.}(2024)\citenamefont {Marini},
  \citenamefont {Calandra},\ and\ \citenamefont {Cudazzo}}]{Marini2024}%
  \BibitemOpen
  \bibfield  {author} {\bibinfo {author} {\bibfnamefont {Giovanni}\
  \bibnamefont {Marini}}, \bibinfo {author} {\bibfnamefont {Matteo}\
  \bibnamefont {Calandra}}, \ and\ \bibinfo {author} {\bibfnamefont
  {Pierluigi}\ \bibnamefont {Cudazzo}},\ }\bibfield  {title} {\enquote
  {\bibinfo {title} {Optical absorption and photoluminescence of single-layer
  boron nitride from a first-principles cumulant approach},}\ }\href {\doibase
  10.1021/acs.nanolett.4c00669} {\bibfield  {journal} {\bibinfo  {journal}
  {Nano Letters}\ }\textbf {\bibinfo {volume} {24}},\ \bibinfo {pages}
  {6017--6022} (\bibinfo {year} {2024})},\ \bibinfo {note} {pMID: 38723148},\
  \Eprint {http://arxiv.org/abs/https://doi.org/10.1021/acs.nanolett.4c00669}
  {https://doi.org/10.1021/acs.nanolett.4c00669} \BibitemShut {NoStop}%
\bibitem [{\citenamefont {Bechstedt}\ \emph {et~al.}(1994)\citenamefont
  {Bechstedt}, \citenamefont {Fiedler}, \citenamefont {Kress},\ and\
  \citenamefont {Del~Sole}}]{Bechstedt1994}%
  \BibitemOpen
  \bibfield  {author} {\bibinfo {author} {\bibfnamefont {F.}~\bibnamefont
  {Bechstedt}}, \bibinfo {author} {\bibfnamefont {M.}~\bibnamefont {Fiedler}},
  \bibinfo {author} {\bibfnamefont {C.}~\bibnamefont {Kress}}, \ and\ \bibinfo
  {author} {\bibfnamefont {R.}~\bibnamefont {Del~Sole}},\ }\bibfield  {title}
  {\enquote {\bibinfo {title} {Dynamical screening and quasiparticle spectral
  functions for nonmetals},}\ }\href {\doibase 10.1103/PhysRevB.49.7357}
  {\bibfield  {journal} {\bibinfo  {journal} {Phys. Rev. B}\ }\textbf {\bibinfo
  {volume} {49}},\ \bibinfo {pages} {7357--7362} (\bibinfo {year}
  {1994})}\BibitemShut {NoStop}%
\bibitem [{\citenamefont {Cudazzo}(2023)}]{Cudazzo2023}%
  \BibitemOpen
  \bibfield  {author} {\bibinfo {author} {\bibfnamefont {Pierluigi}\
  \bibnamefont {Cudazzo}},\ }\bibfield  {title} {\enquote {\bibinfo {title}
  {Dynamical effects on photoluminescence spectra from first principles: A
  many-body green's function approach},}\ }\href {\doibase
  10.1103/PhysRevB.108.165101} {\bibfield  {journal} {\bibinfo  {journal}
  {Phys. Rev. B}\ }\textbf {\bibinfo {volume} {108}},\ \bibinfo {pages}
  {165101} (\bibinfo {year} {2023})}\BibitemShut {NoStop}%
\end{thebibliography}%

\end{document}